# Dislocation-induced Y segregation at basal-prismatic interfaces in Mg

Zhifeng Huang [a, b], Vladyslav Turlo [b], Xin Wang [c], Fei Chen [a, *], Qiang Shen [a], Lianmeng Zhang [a], Irene J. Beyerlein [d], Timothy J. Rupert [b, c, *]

[a] State Key Lab of Advanced Technology for Materials Synthesis and Processing, Wuhan University of Technology, Wuhan 430070, China

[b] Department of Mechanical and Aerospace Engineering, University of California, Irvine, CA 92697, USA

[c] Department of Materials Science and Engineering, University of California, Irvine, CA 92697, USA

[d] Materials Department, University of California, Santa Barbara, CA 93106, USA

* Corresponding author: chenfei027@whut.edu.cn (F. Chen), trupert@uci.edu (T.J. Rupert)

Solute segregation at twin boundaries in Mg has been widely investigated, yet this phenomenon has not been studied at the equally important basal-prismatic interfaces.  To fill this critical gap, this work investigates the segregation behavior of Y at basal-prismatic interfaces with various structures using atomistic simulations.  The calculated interfacial energies show that short coherent interfaces and long semi-coherent interfaces containing disconnections and dislocations are more energetically stable than disordered interfaces, which is supported by our experimental observations.  The segregation energy of Y at these lowest energy basal-prismatic interfaces shows a clear correlation with the atomic hydrostatic stress, highlighting the importance of local extension stresses for segregation.  In addition, sites around dislocations at the semi-coherent basal-prismatic interfaces demonstrate lower segregation energy, indicating that local defects such as interfacial dislocations can further enhance the segregation.  In its entirety, this study indicates that the segregation of solutes can be affected by a number of different aspects of the local structure at complex interfaces in Mg alloys.

# 1. Introduction

Mg and Mg-rich alloys are promising structural materials for transportation applications because of their low density and high specific strength, which would significantly increase efficiency and reduce environmental impact [1]. Generally, compared to pure Mg, Mg alloys exhibit enhanced ductility and strength due to the effect of solutes on the stability and mobility of crystalline defects [2, 3]. Specifically, twin boundaries play an essential role in Mg and Mg alloys, as twinning is a major deformation mode due to the limited number of easily activated dislocation slip systems [4-6]. In addition, twins can be weak points in some cases, such as when these features act as crack nucleation sites under tension because of local strain concentration [7]. Solute segregation to twin boundary sites has been experimentally observed to be widespread in Mg alloys, and this phenomenon can tailor interfacial energy and properties [3, 8-16]. For example, Nie and coworkers found that the chemically periodic segregation of solutes such as Gd [8, 9], Zn [10], Ag [12], and Nd [12] at twin boundaries can provide strong pinning effects on twin boundary migration. Tsuru et al. reported that solutes such as Y, Zn, Ag, and Mn that segregate to twin boundaries can enhance interfacial cohesion and prevent crack propagation along with the interfaces [3, 17, 18]. The segregation behavior of various solutes at twin boundaries has been systematically investigated on the basis of density functional theory (DFT) calculations [14, 15], and results reveal that larger solutes prefer to occupy sites that are under a local tensile stress whereas smaller solutes occupy sites that are under compression at twin boundaries to lower the interfacial energy.

The vast majority of prior work in this area has considered solute segregation at the symmetric $\{10\bar{1}1\}$, $\{10\bar{1}2\}$ and $\{10\bar{1}3\}$ coherent twin boundaries (TBs) [3, 14-16, 19], while this phenomenon has not been studied at the equally important basal-prismatic/prismatic-basal (BP/PB) interfaces commonly observed experimentally [15, 20-25]. BP/PB interfaces are asymmetric boundaries attached to $\{10\bar{1}2\}$ coherent TBs. The BP/PB interfaces are needed for the migration of TBs, as they allow for structures that vary parallel to the TB plane normal or to create connected boundaries at twin tips inside of grains. The choice of BP or PB designation depends on which lattice plane is oriented normal to the interface inside the twin and matrix, so we simplify the terminology to just "BP" for the rest of this paper. Previous experimental observations have shown that there are two configurations of BP interfaces: (1) short, coherent interfaces with a simple structure containing the same number of atomic layer of the basal plane and the prismatic plane [15], and (2) long, semi-coherent interfaces with different numbers of atomic layers of the basal plane and the prismatic plane which also contain a series of disconnections and dislocations [20-25]. BP interfaces are necessary features of growing twins and intimately involved in twinning and detwinning deformation. For example, the long BP interfaces can be relatively immobile and have to decay into more mobile and serrated short BP interfaces prior to migration under loading [21]. While Kumar et al. [15] studied the segregation of solutes at a short coherent BP interface, the segregation behavior of solutes has not been reported for the other BP interfaces. In addition, prior studies have primarily focused on the investigation of the interfacial energies and atomic structures of the coherent BP interfaces [15] and the lowest misfit 15:16 BP interfaces [26, 27], with these numbers denoting the number of basal and prismatic lattice planes, respectively. Here, we refer to the number of terminating basal planes with the variable *k*.

In our study, we first investigate the atomic structures and stability of various BP interfaces by combining transmission electron microscopy (TEM) characterization of real twin structures with molecular dynamics (MD) simulations. The results show that interfacial energies of the semi-coherent $k$:($k$+1) ($k \geq 15$) BP interfaces are much lower than the disorder $k$:($k$+1) ($k < 15$) BP interfaces and this energy decreases with increasing $k$ when $k \geq 15$. The relatively low energy options are fully coherent interfaces and semi-coherent interfaces, which agree with the two types of interfacial configurations that were observed in TEM. Next, we calculate the segregation energy of Y at these lower energy interfaces, with Y chosen as the solute because it is a common alloying element for Mg and has been previously shown to segregate to twin and other grain boundaries [15, 28, 29]. Our results show that Y preferentially segregates to the sites with large atomic hydrostatic stress, with dislocations at the $k$:($k$+1) BP interfaces being sites for strong segregation.

## 2. Methods

### 2.1 Experimental materials processing and characterization

Hot rolled Mg-1 wt.% Y was recrystallization annealed at 400 °C for 10 min to achieve a nearly equiaxed and twin-free microstructure with an average grain size of ~22 μm. The alloy plate was sectioned into 4 × 4 × 6 mm cuboids using electrical discharge machining for subsequent compression tests. The samples were quasi-statically compressed at room temperature along the rolling direction to a true strain of 2% to generate $\{10\bar{1}2\}$ tensile twins. The deformed samples were then sectioned into ~ 1 mm thick thin foils parallel to the rolling direction and the normal direction. The foils were mechanically polished using diamond suspensions, followed by chemical etching using a 10% Nital to remove any damaged surface layers. A Gatan PIPS system was used to prepare electron transparent specimens, which were then investigated with a JEOL JEM-2100F TEM operating at 200 kV.

**2.2 Computational methodology**

Atomistic simulations are performed by using the large-scale Atomic/Molecular Massively Parallel Simulator (LAMMPS) package [30] with the modified embedded atom method (MEAM) Mg-Y potential, developed by Kim et al. [31], which has been verified to agree well with experimental measurements and density functional theory (DFT) calculations for physical properties such as surface energy, elastic constants, stacking fault energies, and twin boundary energies [28, 31]. In addition, the potential has been verified to agree well with experimental observations of Y segregation to grain boundaries in Mg [31], which is important for this specific study. BP interfaces with ratios of 1:1 and $k$:($k$+1) were created by combining two cells with the appropriate number of lattice planes. We focus our attention on these BP interfaces because our experimental observations show that two types of BP interfaces exist: (1) 1:1 BP interfaces and (2) $k$:($k$+1) BP interfaces, as shown in Fig. 1. For the 1:1 and $k$:($k$+1) BP interfaces when $k$ < 15, the basal lattice is under tension and the prismatic lattice is under compression. In contrast, for $k \geq 15$, the basal lattice is under compression while the prismatic lattice is under tension. Each lattice is pre-strained to match one another and then relaxed, before finally joining the two simulation cells to form the PB boundary. The boundary energy is obtained by calculating the energy difference between the model containing the boundary and the corresponding perfect crystal. Atomic snapshots are visualized using the Open Visualization Tool (OVITO) [32]. The segregation behavior of Y is investigated at various BP interfaces by replacing a Mg atom within a 2 nm centered on the interface by a Y atom. The segregation energy ($E_{seg}$) is calculated to determine the segregation ability according to [14, 15, 33]:

$$E_{seg} = (E_{BP+Y} - E_{BP}) - (E_{bulk+Y} - E_{bulk}) \qquad (1)$$

where $E_{BP+Y}$ and $E_{BP}$ are the total energies of the supercell with and without Y, while $E_{bulk+Y}$ and $E_{bulk}$ are the total energies of bulk supercells with and without Y.

To verify the reliability of segregation trends calculated by the MEAM potential, density functional theory (DFT) calculations of segregation to the coherent 1:1 BP interface and twin boundary were performed with the Vienna ab-initio simulation package (VASP) using the projector augmented wave (PAW) pseudo-potentials [34] and the Perdew-Burke-Ernzerhof exchange-correlation generalized gradient approximation (GGA) functional [35]. Supercells of the 1:1 BP interface and the {10-12} coherent twin boundary were built by doubling the sizes in *x* and *y* directions of the 1:1 BP and TB supercell models reported by Kumar et al. [15]. The sites within 8 atomic layers centered on the interfaces were considered as possible segregation sites in the DFT calculations. We note that while this is possible for the short interfacial segments, many of the longer interfaces would require computational cells which are too large for DFT calculations.

## 3. Results and discussion

### 3.1 Structures of BP interfaces from experiments

To provide a baseline expectation for the types of BP interfaces to be investigated, we first performed an experimental study of the BP boundary structure. Fig. 1 shows the morphology of a $\{10\bar{1}2\}$ twin boundary and a twin tip, where the ratio of the atomic layer of the basal plane and prismatic plane are marked on the BP interfaces (blue).  The length and atomic layer ratio of the BP interfaces varies from boundary-to-boundary, in agreement with previous observations [20-25].  Short BP interfaces have the same number of terminating basal and prismatic planes (1:1 ratio), while the long BP interfaces contain one less atomic layer on the basal plane than the prismatic plane ($k$:($k$+1) ratio).  We therefore focus our attention on these interfaces in our atomistic modeling tasks.

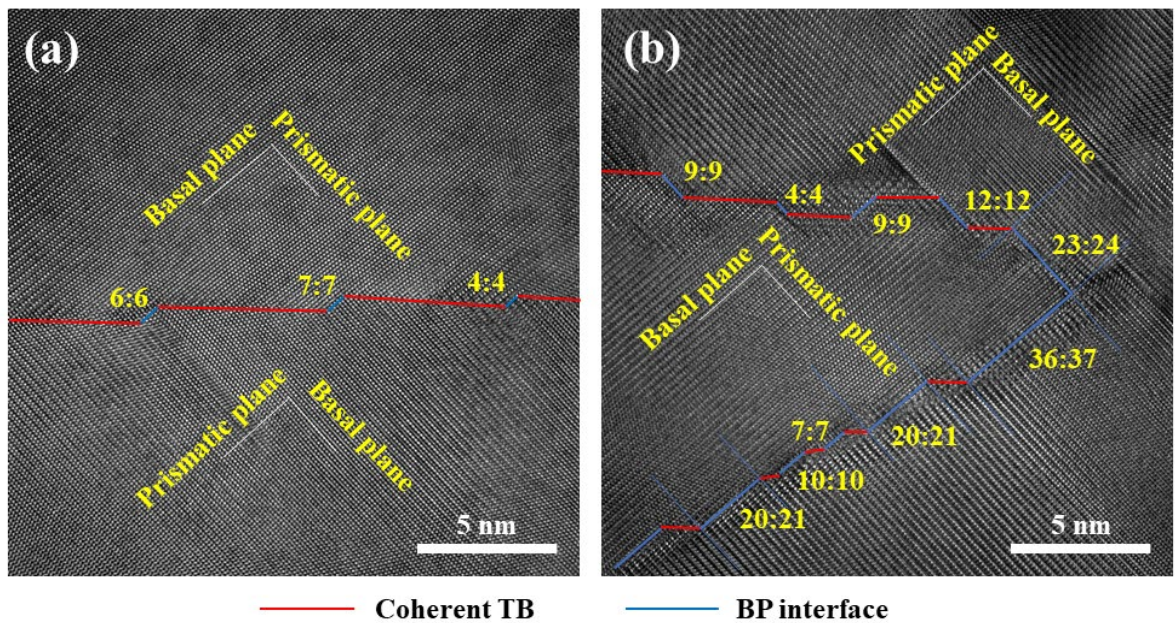

Fig. 1 TEM images of a $\{10\bar{1}2\}$ (a) twin boundary and (b) twin tip in deformed Mg-1 wt.% Y.

### 3.2 Structures and energies of BP interfaces from atomistic simulations

The interfacial energies of BP interfaces are calculated and compared with previous DFT and molecular dynamics (MD) calculations and the results are summarized in Table 1 [15, 26, 27, 36].  The energy of the $\{10\bar{1}2\}$ coherent TB is also shown for comparison purposes. The absolute energy values obtained from different simulation methods or potentials are different, but the trend is the same.  The 1:1 BP interface has the lowest energy, followed by the $\{10\bar{1}2\}$ coherent TB and the 15:16 BP interface, which demonstrates the reliability of our calculations. Considering the misfit strain is ~6.7% for the experimentally observed short 1:1 BP interfaces and that this term is the factor that can lead to the formation of the semi-coherent boundaries, we have only calculated interfacial energies of $k$:($k$+1) BP interfaces with misfits smaller than 6.7%, as shown in Fig. 2(a). This figure shows that the interfacial energies of BP interfaces with $k$ smaller than 15 are > 450 mJ/m$^2$, even when the misfit decreases to 0.2% for the 14:15 BP interface. When $k$ is greater than or equal to 15, the interfacial energies are much lower and this energy decreases with increasing $k$. The results indicate that the $k$:($k$+1) BP interfaces are possible options when $k \geq 15$, which explains why the experimental data in Fig. 1(b) demonstrated 20:21, 23:24, and 36:37 BP interfaces. Fig. 2(b) shows the atomic structures of a 1:1 BP interface containing three lattice periods in the Z-direction and several $k$:($k$+1) BP interfaces ($k$ = 7, 14, 15 and 19). The atomic structure of the 1:1 interface is fully coherent and consistent with prior DFT calculations [15]. The atomic structures of the 7:8 and 14:15 BP interfaces are primarily disordered, with atoms at the interfaces not relaxing to low energy structures. In contrast, the atomic structures of the 15:16 and 19:20 BP interfaces are highly ordered, containing a series of disconnections and dislocations to maintain coherency at the interface. In general, even the lower energy semi-coherent $k$:($k$+1) have a higher interfacial energy than the coherent 1:1 interfaces. However, but it is important to notice that they also have lower misfit strains. This suggest that 1:1 boundaries are preferred for short segments while $k$:($k$+1) boundaries are preferred for longer segments, which again is consistent with experimental observations.

**Table 1** Comparison of interfacial energies (mJ/m$^2$) for important BP interfaces and the $\{10\bar{1}2\}$ TB calculated in this study, compared to previous work from the literature.

| Interface Type | This Study | DFT (literature) | EAM (literature) |
|---|---|---|---|
| 1:1 BP | 117 | 101 [15], 101-109 [19] | |
| 15:16 BP | 187 | | 164 [26, 27] |
| {10-12} TB | 143 | 135 [15] | 124 [36] |

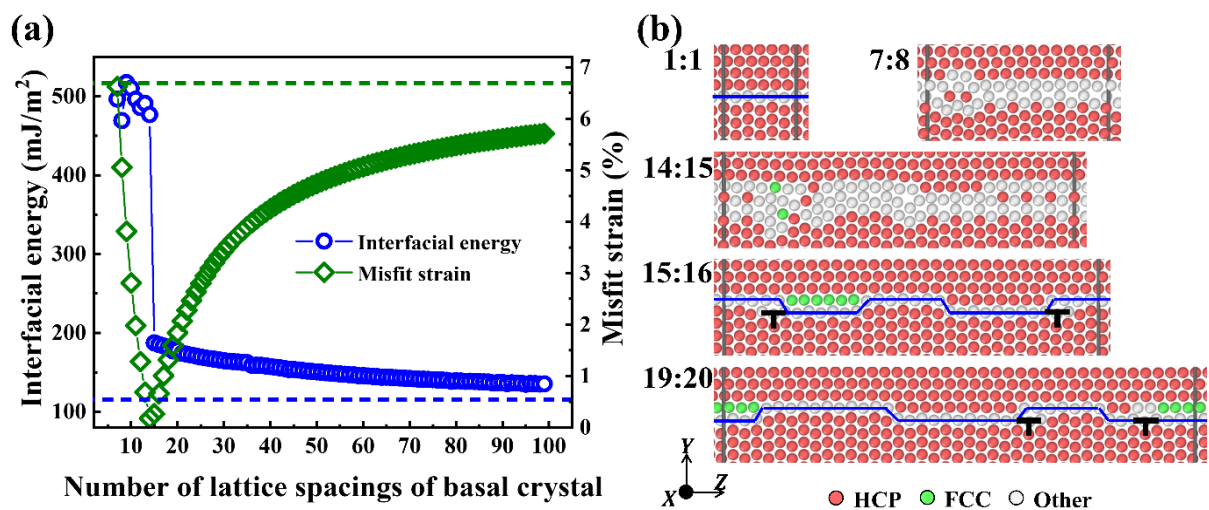

**Fig. 2** (a) Calculated interfacial energy and misfit strains for 1:1 (horizontal dashed lines) and $k$:($k$+1) BP interfaces (data points). (b) Atomic structures of several representative BP interfaces, with examples of interfacial dislocations marked.

### 3.3 Segregation energies of Y at BP interfaces

With the baseline structures found, we turn our attention to study the segregation ability of Y at the 1:1, 15:16, and 19:20 BP interfaces. Previous studies have suggested that the segregation ability of a solute is mainly dominated by local distortion (atomic volume or local elastic strain) at the interfaces [10, 12, 14]. Therefore, the segregation energies of Y for different sites are plotted as a function of atomic volume or hydrostatic stress at possible segregation sites in Fig. 3. We also plot the segregation energies for Y at different sites near the $\{10\bar{1}2\}$ coherent TB for comparison. Fig. 3(a) and (b) show the segregation energy of Y plotted against change of atomic volume, as calculated by both DFT and MEAM. The change of atomic volume is calculated as $(V_{interface} - V_{perfect})/V_{perfect}$, where $V_{interface}$ and $V_{perfect}$ are the atomic volume of a Mg atom at or close to the interface and the atomic volume of a Mg atom in the perfect Mg crystal, respectively. Although the energy values calculated by DFT and MEAM are different in absolute value, the same trend of segregation behavior is observed. The segregation energy decreases with increasing change of atomic volume in both DFT and MEAM calculations, demonstrating that the ability to accommodate the larger Y atom is key. The observation of similar trends provides additional evidence of the reliability of the MEAM potential for simulating Y segregation to the interfaces. The relationship of the atomic hydrostatic stress and change of atomic volume calculated by the MEAM potential is also plotted in Fig. 3(a) and (b). Positive and negative values of the atomic hydrostatic stress indicate sites under hydrostatic tension and compression, respectively. A positive correlation between the atomic hydrostatic stress and the change of atomic volume is clearly observed, demonstrating that Y prefers to segregate to the sites with large positive hydrostatic stresses. The relationships between segregation energy and atomic hydrostatic stress for the semi-coherent 15:16 and 19:20 BP interfaces calculated by the MEAM potential are plotted in Fig. 3(c) and (d). Most of the data show a similar correlation between segregation energy and atomic hydrostatic stress, although several data points deviate from this trend and appear below the rest. Hypothesizing that these deviations occur when significant boundary structural changes occur with doping, we also calculated the segregation energy with all atomic sites fixed ($E_{seg}^{fixed}$) for the 15:16 and 19:20 BP interfaces, as shown in Fig. 3(c) and (d). The results show that in the absence of any structural relaxation, the segregation ability of Y at a site is only dependent on the hydrostatic stress of the original site and the data converge to a single line.

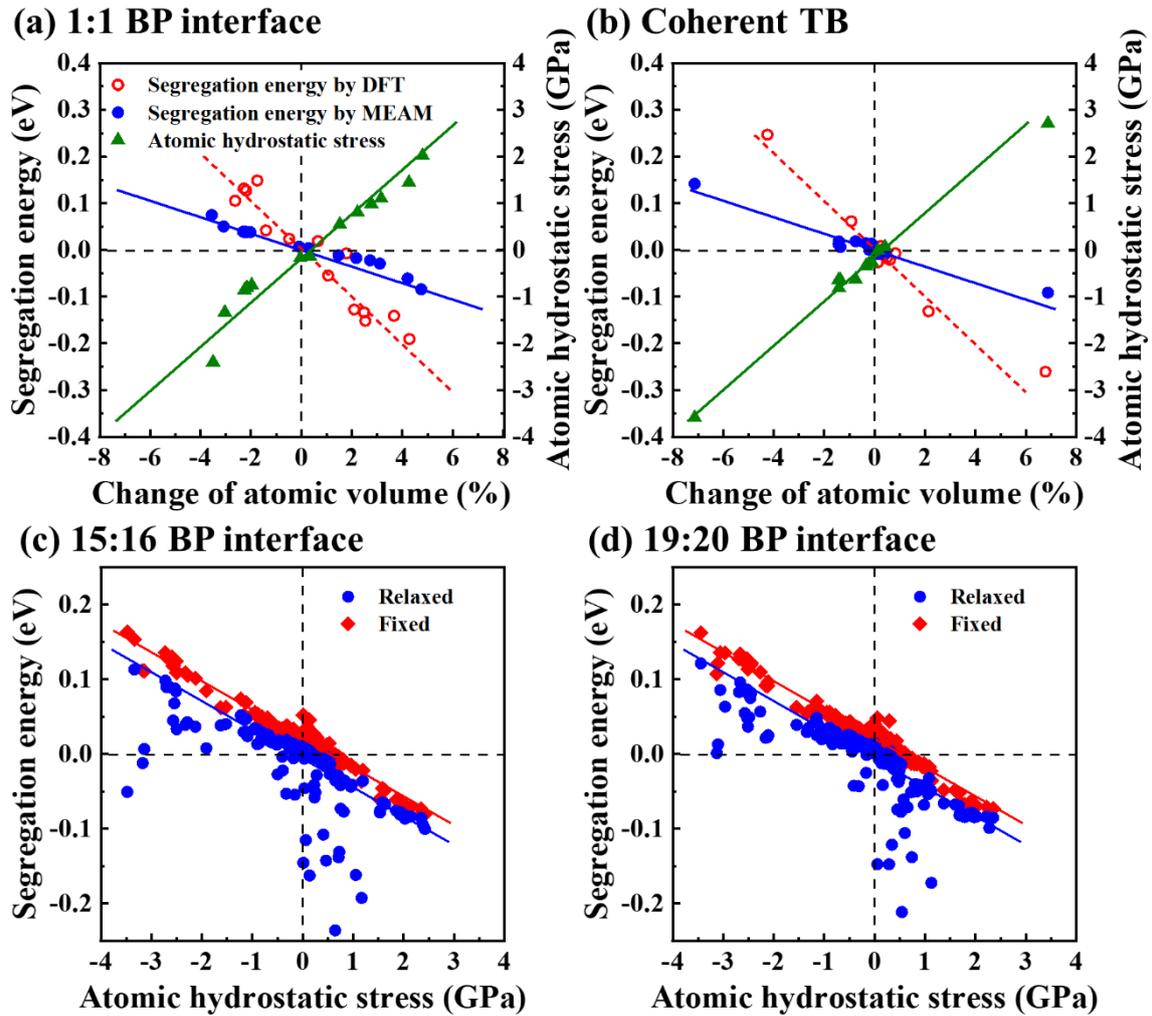

Fig. 3 Segregation energy of Y and atomic hydrostatic stress as a function of the change of atomic volume (%) across (a) the 1:1 BP interface and (b) the coherent TB. Both DFT and MEAM calculations are presenting in (a) and (b). Segregation energy of Y as a function of atomic hydrostatic stress for (c) the 15:16 and (d) the 19:20 BP interfaces, as calculated using the MEAM potential.

The disconnections and dislocations shown in Fig. 2(b) are the likely sites for noticeable relaxation during doping, so we investigate these defects further. Figs. 4(a) and (b) show the atomic distributions of hydrostatic stress and relaxed segregation energies ($E_{seg}^{relaxed}$) of Y at different sites across the 15:16 and 19:20 BP interfaces. It is clear that the regions of large hydrostatic stresses are mainly located at the interface. Generally, the sites with large positive values of hydrostatic stress are the sites with large negative values of $E_{seg}^{relaxed}$, with periodic locations along the BP interface being likely segregation sites. However, some strong segregation sites exist at locations where there is an absence of an elevated hydrostatic stress, most notably near the labeled grain boundary dislocations. To highlight these defects, we also plot the shear stress along the *XZ* direction ($\tau_{xz}$) and the difference of the relaxed segregation energy and the fixed segregation energy ($E_{seg}^{relaxed} - E_{seg}^{fixed}$) in Fig. 4(c) and (d). The two opposite $\tau_{xz}$ stress fields around the two dislocations indicating that these are mixed dislocations with a pair of disclinations in the opposite directions. This finding is similar to a previous observation for the 15:16 BP interface [26, 27], with a difference in the location of dislocations likely caused by use of different interatomic potentials. It is clear to see that the sites with large differences in relaxed and fixed segregation energies are indeed the sites around the interfacial dislocations. Fig. 4(e) shows the relaxation displacement magnitudes of atoms near the interfacial dislocations and disconnections of BP interfaces after relaxation. These enlarged views in Fig. 4(e) correspond to the red rectangles in Fig. 4(d), and the Y atoms are marked with light blue circles in Fig. 4(d) and as blue atoms in Fig. 4(e). The atomic displacement magnitudes near the Y segregation sites around the dislocations are obviously larger than those around the disconnections. The local structural changes suggest that interfacial dislocations can more easily free up enough space for Y at BP interfaces, highlighting the sites which showed deviations below the energy-stress trend line in Figs. 3(c) and (d).  Notably, this finding demonstrates that interfacial defects can further increase the ability of an interface to attract dopant elements.

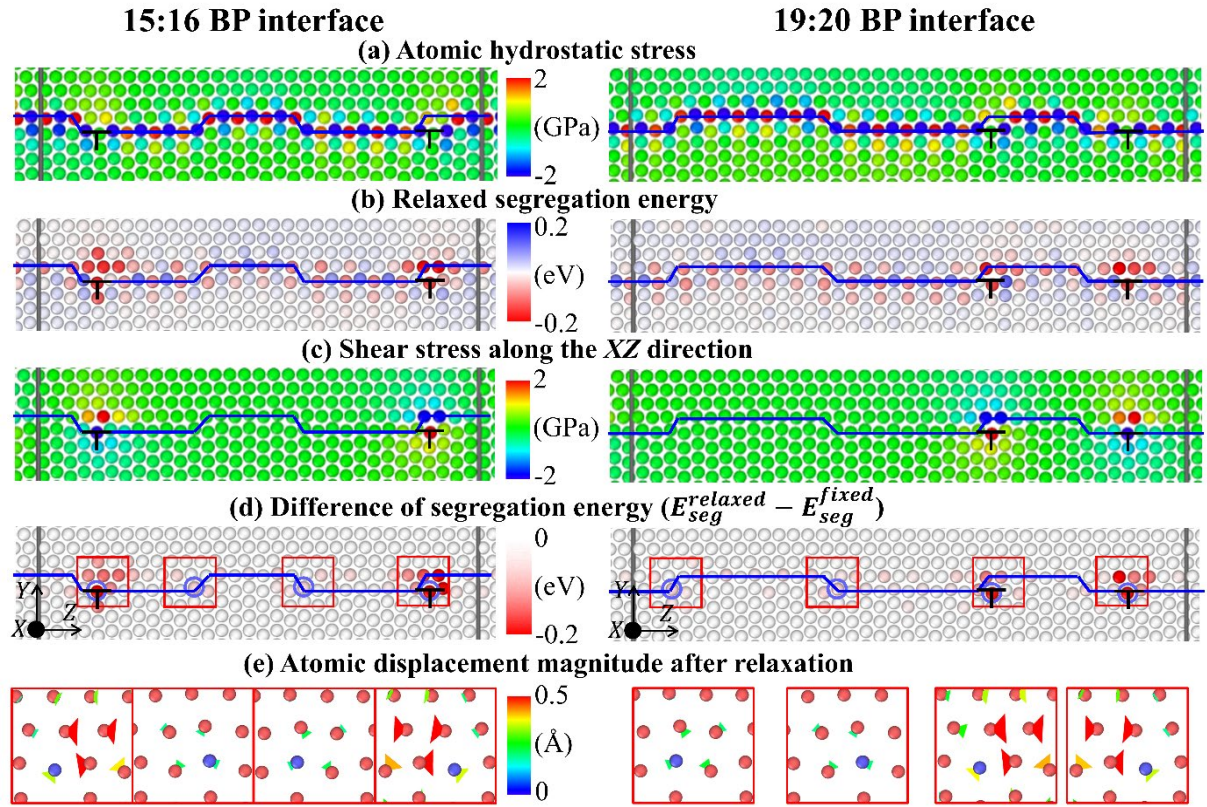

Fig. 4 The atomic distributions of (a) hydrostatic stress, (b) relaxed segregation energy ($E_{seg}^{relaxed}$), (c) shear stress along the *XZ* direction ($\tau_{xz}$), and (d) difference of segregation energy ($E_{seg}^{relaxed} - E_{seg}^{fixed}$).  (e) Local displacement magnitudes near dislocations and disconnections after relaxation for the 15:16 and 19:20 BP interfaces.  The blue and red atoms in (e) denote Y and Mg atoms, respectively.

Previous studies have mainly considered the segregation of Y to the coherent TB and coherent BP interface [14, 15].  However, referring back to Fig. 3, the segregation energies for Y around dislocations at the semi-coherent BP interfaces are significantly lower than the values found at coherent TB and coherent BP interfaces, indicating that semi-coherent BP interfaces with defects should be the preferred sites for Y clustering in real materials.  Rare earth (RE) element clusters at interfaces have indeed been reported in other Mg-RE alloys [9, 13], although not specifically for Y to date.  For example, Luo et al. [13] reported Gd clusters at dislocations along grain boundaries during high resolution TEM experiments.  In addition, these authors measured a significant improvement in alloy strength due to the pinning effect of Gd clusters at the grain boundaries, as well as an improvement in ductility, which was mainly attributed to the inhibition of premature failure due to Gd segregation at twin boundaries and non-basal slip.  Therefore, the results presented here suggest that increased attention should be paid to investigating the structure and behavior of BP interfaces in the future, in order to comprehensively understand the effect of solute segregation on microstructure evolution and mechanical properties of Mg alloys.

## 4. Conclusion

This work systematically reports on the segregation behavior of Y at a series of BP interfaces.  We first investigate the atomic structures and energetics of various BP interfaces by combining TEM characterization and atomistic simulations.  Our results show that interfacial energies of the coherent 1:1 and semi-coherent $k$:($k$+1) (with $k \geq 15$) BP interfaces are much lower than the disordered $k$:($k$+1) (with $k < 15$) BP interfaces, which is consistent with the interfacial configurations observed in experiments.  Next, the segregation energies of Y at various BP interfaces are calculated and show a clear negative correlation between the segregation energy and the atomic hydrostatic stress.  In addition, dislocations at BP interfaces can further enhance Y segregation.  In general, this work shows that interfacial segregation can be affected by the local structure and defect content at complex interfaces.

**CRediT authorship contribution statement**

Z.H. and V.T. Conceptualization, Methodology, Investigation, Writing - original draft, Writing - review & editing; X.W. Experiment, TEM observation, Writing - review & editing; F.C., Q.S., L.Z., I.J.B., and T.J.R. Conceptualization, Writing - review & editing, Supervision, Project administration, Funding acquisition.

**Declaration of Competing Interest**

The authors declare that they have no known competing financial interests or personal relationships that could have appeared to influence the work reported in this paper.

**Data availability**

The raw/processed data required to reproduce these findings cannot be shared at this time as the data also forms part of an ongoing study.

**Acknowledgments:**

T.J.R. acknowledges discretionary support from the Henry Samueli School of Engineering. Z.H., F.C., Q.S., and L.Z. also acknowledge the Fundamental Research Funds for the Central Universities in China.